\documentclass[10pt,aps,prb,twocolumn,superscriptaddress,floatfix]{revtex4-2}
\usepackage[margin= 0.6in,top= 18.0mm]{geometry}
\usepackage{amsmath,amssymb,bm} 
\usepackage{graphicx}
\usepackage{epstopdf}
\usepackage{latexsym}
\usepackage[dvipsnames]{xcolor}
\usepackage{braket}
\usepackage{float}
\usepackage{placeins}
\usepackage[normalem]{ulem}
\usepackage{comment}
\usepackage{mathtools}
\usepackage{array}
\usepackage{tabu}
\usepackage{multirow}
\usepackage[inline]{enumitem}
\usepackage[normalem]{ulem}
\usepackage{natbib}  
\usepackage[version=4]{mhchem}
\usepackage{csquotes}
\usepackage[colorlinks=true, linkcolor=blue, citecolor=blue, urlcolor=blue]{hyperref}
\usepackage{cleveref}

\begin{document}
\title{Interplay of Excitonic Charge Density Wave and Superconductivity in Transition Metal Dichalcogenides}
\author{Ayussh A. Patel}
\affiliation{Department of Physical Sciences, Indian Institute of Science Education and Research Kolkata, Mohanpur 741246, India}
\author{Amit Ghosal}
\affiliation{Department of Physical Sciences, Indian Institute of Science Education and Research Kolkata, Mohanpur 741246, India}

\begin{abstract}
Motivated by the unique characteristics of the phase diagram of
$1T$-\ce{TiSe_2} we investigated the
complex interplay of excitonic charge density wave (CDW) and superconductivity (SC) on a two-dimensional triangular lattice, each site accommodating two orbitals. In response to various tuning parameters, such as intercalation, pressure, substitution, and gating, these materials exhibit a generic and gradual waning of CDW, followed by the emergence of SC.
Intriguingly, the nature of CDW changes from commensurate to incommensurate with the appearance of SC.
Setting up a minimal model based on observed excitonic CDW and electron and hole pockets in the Fermi surface, and analyzing it within a simple mean-field framework, we can comprehend some salient experimental features. 
We also found that the qualitative phase diagram remains insensitive to details of the band structure. These results offer crucial insights into the nature of the interplay between CDW and SC in transition-metal dichalcogenides.
\end{abstract}

\maketitle

\section{Introduction}

The interplay between symmetry-broken orders in the ground states (GS) is the hallmark of correlated quantum matter. These include Heavy Fermion Systems~\cite{Coleman2007HeavyFermions}, High temperature superconductor (SC) to manganites featuring colossal magnetoresistance~\cite{Dagotto2005Complexity}, Pnictides~\cite{Paglione2010}, single-phase multiferroics~\cite{Eerenstein2006}, and, more recently Moire superlattices~\cite{Caoetal2018,CaoUnconventional2018} which have been revealing an increasingly rich landscape of competing and collaborating phases by tuning relevant parameters.
It is the complex teamwork of various ordered phases in different parameter regimes that ascribes the richness of correlated matter. Among these, the competition of superconductivity (SC) and charge density wave (CDW) has a long history. It has been studied on different systems, unraveling diverse facets of their interrelationship. In recent times, transition metal dichalcogenides (TMDs) offer a new angle~\cite{Rossnagel2011CDW,Hwang2024CDW,KLEMM201586,Xu2021CDW} to explore such an interplay.


A prototype example of TMDs that has gone through a variety of experimental scanners is $1T$-\ce{TiSe_2}. The monolayer form develops a $2\times2$ commensurate CDW (C-CDW) at $T_c\approx232K$~\cite{Chen2015,doi:10.1021/acsnano.5b06727,doi:10.1021/acs.nanolett.6b02710}. The modulation pattern is known to generate a hexagonal $3\bf{Q}$ spatial structure in the GS~\cite{PhysRevB.14.4321}. However, a stripe phase $1{\bf Q}$ has also been realized \cite{PhysRevLett.131.196401,PhysRevLett.118.017002,PhysRevB.110.165156}.
The normal state of this materials is believed to be a semi-metal~\cite{PhysRevLett.99.027404} or a semiconductor with a small indirect band gap~\cite{PhysRevLett.101.237602} owing to contributions from electron pocket at $M$-point and hole pockets at $\Gamma$-point of the fermi-surface (FS) also hinting at the possible excitonic origins of the charge order in the system~\cite{doi:10.1126/science.aam6432,Monney_2010,PhysRevB.103.125430}. 
$1T$-\ce{TiSe_2} develops superconductivity with intercalation~\cite{Morosan2006}, substitution~\cite{PhysRevB.110.165156} or by tuning carrier density~\cite{Li2016} and pressure~\cite{PhysRevLett.103.236401}. The emergent SC is generally conventional~\footnote{if tuning parameter is pressure, the conventional SC is realized only for low pressure~\cite{https://doi.org/10.1002/smll.202402749}.}.
It is the universality of the broad topology of the phase diagram featuring SC and CDW under a wide range of tuning probes (such as intercalation, doping, and pressure) that attracted significant research attention.
A microscopic look into these materials using scanning tunneling spectroscopy (STS) probes identified domains of CDW puddles~\cite{PhysRevLett.118.106405,Joe2014} above the SC-dome in the phase diagram, which suggested an incommensuration of CDW (I-CDW) as a precursor to the onset of SC.
\begin{figure}[b]
    \centering
    \includegraphics[width=0.5\textwidth]{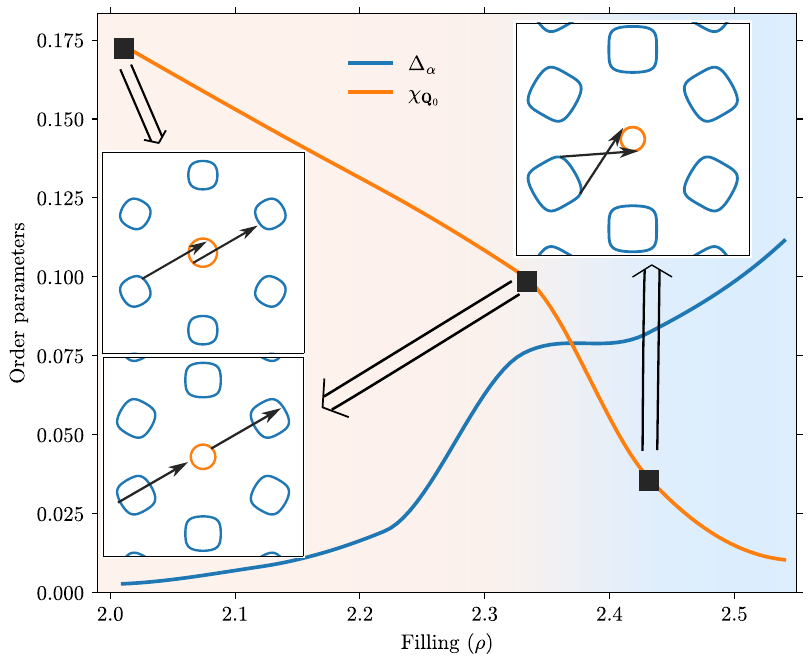}
    \caption{Energy-minimized phase diagram obtained by changing electronic density. The system undergoes a cross-over from C-CDW to NC-CDW order. Insets show the schematics of the relevant scattering in the system. Refer Fig~\ref{fig:fig5} for details.}
    \label{fig:fig1} 
\end{figure}

Different theoretical directions have been pursued to comprehend this large body of intriguing experimental signatures. These include
a first principle density functional calculations~\cite{PhysRevLett.106.196406,Hu_2021,PhysRevB.94.214507,PhysRevB.104.184506}, Ginsburg-Landau theories~\cite{PhysRevB.99.121108,PhysRevB.101.195424}, mean field calculations on microscopic models of electronic interactions~\cite{PhysRevB.81.165109,PhysRevLett.121.226602,PhysRevB.98.104206} and symmetry analysis~\cite{Kim2024ChiralCDW}. Although these studies unveiled different phenomenological aspects, a minimal solvable model to describe the key signature of these materials, namely, the C-CDW and I-CDW phases along with the SC dome in the phase diagram, which are insensitive to the choice of tuning parameters, remains elusive.
Our present study addresses precisely these points. The key findings are summarized in Fig.~\ref{fig:fig1} displaying the phase diagram and is elaborated in the following.


Fig.~\ref{fig:fig1} illustrates a portion of the phase diagram at zero temperature ($T=0$) of a generic TMD compound, considering the charge carrier density, $\rho$, as the tuning parameter.
At half-filling ($\rho=2.0$) a C-CDW constitutes the GS. 
As $\rho$ increases gradually, the CDW order parameter suffers a strong reduction (orange trace) and simultaneous nucleation of SC (blue trace). Interestingly, the system maintains the same C-CDW GS up to $\rho=2.33$. For $2.33 < \rho < 2.44$, although a commensurate CDW continues as the GS, the ordering wave-vector becomes $\rho$ dependent and splits into two.
This is indicative of a changeover of the GS that supports a non-commensurate CDW phase (NC-CDW) instead of C-CDW. This is illustrated in the insets of Fig.~\ref{fig:fig1} by highlighting the relevant scatterings in terms of arrows on the Fermi surface (FS). This will be discussed in detail in Sec.~\ref{sec:sec3c}. 
Interestingly, this changeover occurs in the same window of $\rho$ where the CDW order parameter suffers a strong reduction and simultaneous rise of SC order. This is tantalizingly similar to the experimental phase diagram~\cite{Morosan2006,PhysRevB.110.165156,Li2016,PhysRevLett.103.236401}.

The plan for the rest of the paper is as follows: In Sec.~\ref{sec:2}, we introduce our toy model to describe the generic features of TMDs and also describe the methodology of our mean-field calculations. 
In Sec.~\ref{sec:sec3}, we describe our results
focusing in individual subsections on the phase diagram, the evolution of the spectral function and the evolution of the associated FS, as well as the density of states (DOS). In discussing these results, we will focus on the relevance of our findings in addressing key experimental features, namely (i) excitonic nature of CDW, (ii) NC-CDW as precursor to SC, and (iii) electronic nature of the cross-over.
Finally, we discuss and conclude in Sec.~\ref{sec:sec4}.
We have also included some relevant details in the Appendix.

\section{Model and Methods}
\label{sec:2}

Our minimal toy model consists of a one-particle band Hamiltonian, ${\cal H}_0$, and the interaction term that generates inter-band CDW, as well as intra-band SC. Choices of such band-dependence of the symmetry breaking ordered ground state is purely phenomenological--while the experiments on $1T$-$\ce{TiSe_{2}}$ hint towards the exitonic nature of the CDW   ordering~\cite{doi:10.1126/science.aam6432,Monney_2010,PhysRevB.103.125430}, the SC is usually considered to be conventional~\cite{https://doi.org/10.1002/smll.202402749}. This motivates us to consider the conventional BCS pairing for SC, keeping in mind the need for a minimal yet reasonable description. In the following, we discuss the proposed ${\cal H}_0$ and ${\cal H}_{\rm int}$ separately.

\subsection{Band Hamiltonian} 

\begin{figure}[t]
    \centering
    \includegraphics[width=0.5\textwidth]{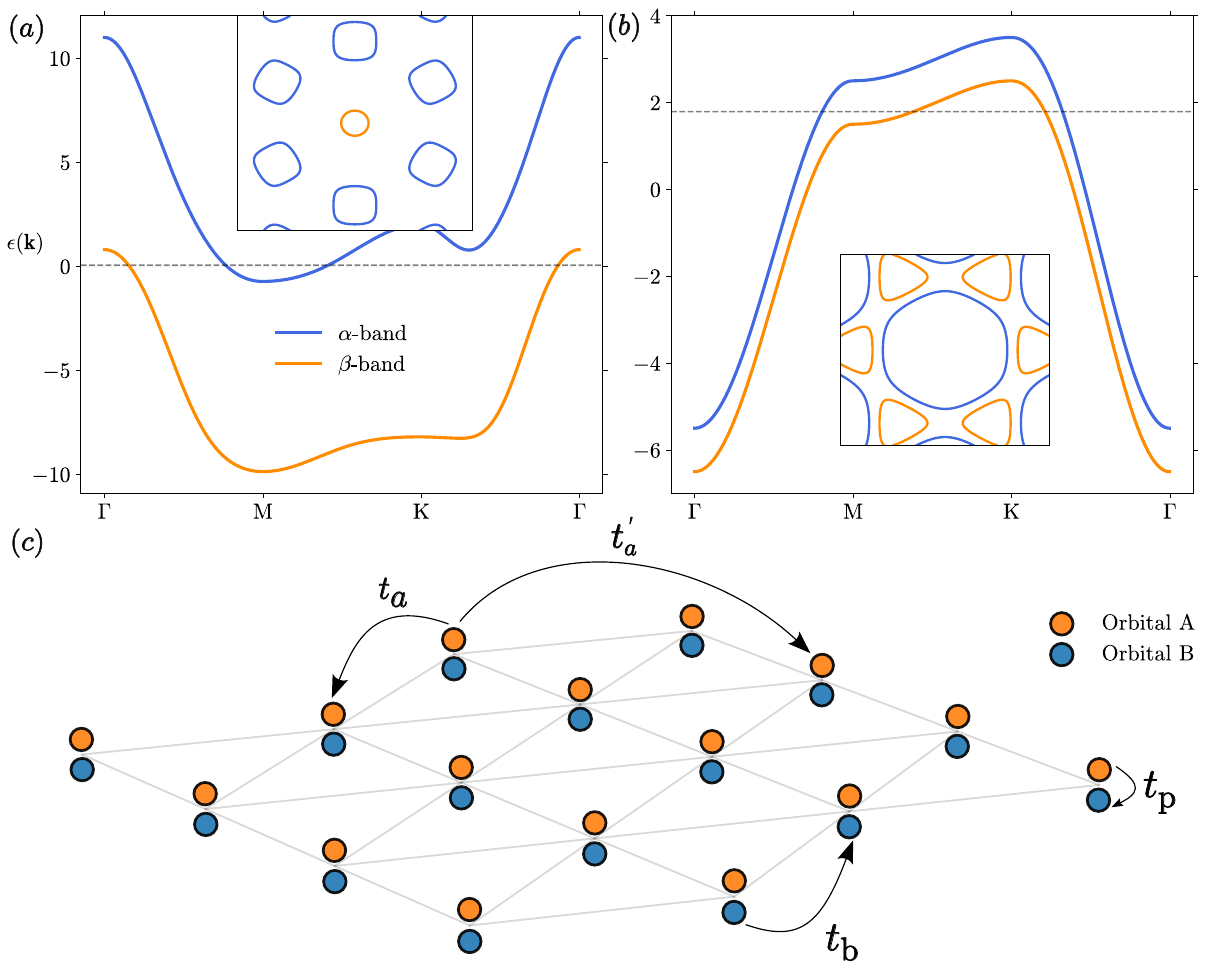}
    \caption{(a) Primary model: On-site energies $\epsilon_A=\epsilon_B =-2.5$, nearest neighbor hopping $t_a=t_b=1.0$, Next nearest neighbor hopping only in A-orbital  $t_a^{'}=0.8$, on-site inter-orbital hopping is $t_p=4.5$  (b) and for Secondary model: $\epsilon_{0}^{a} =0.0,t_a=t_b=-1.0,t_a^{'}=0.0,t_p=0.5$ (inset of both figure show the geometry of the fermi surface along the dashed black line). (c) The underlying microscopic tight-binding model which, upon diagonalization, generates the bands in (a) and (b).}
    \label{fig:fig2}
\end{figure}

A quantitative description of the electronic sector in $1T$-\ce{TiSe_{2}} usually requires five ${\rm 3d}$ orbitals of the \ce{Ti} atom and three $4p$ orbitals for each of the two \ce{Se} atoms, totaling $11$ orbitals~\cite{PhysRevLett.106.106404}. However, to capture the essential multi-orbital physics, we resort to a minimal description considering a two-orbital model with inter-orbital coupling, which upon diagonalization yields a $2$-band Hamiltonian: 
${\cal H}_{\rm band}=\sum_{k,\sigma}\epsilon_k^{\alpha} \alpha_{k\sigma}^\dagger \alpha_{k\sigma} + \epsilon_k^{\beta} \beta_{k\sigma}^\dagger \beta_{k\sigma}$. Here, $\alpha_{k\sigma}^\dagger$ ($\alpha_{k\sigma}$) creates (annihilates) an electron in band $\alpha$,
and similarly $\beta_{k\sigma}^\dagger$ ($\beta_{k\sigma}$) are defined on the $\beta$-band. The chosen band-structures, i.e., the $k$ dependence of $\epsilon_k^\alpha$ and  $\epsilon_k^\beta$, are presented in Fig.~\ref{fig:fig2}(a) along the high-symmetry lines (the direction $\Gamma$-$ K$-$M$-$\Gamma$ in the ${\bf k}$-space) of the first Brillouin zone (1BZ). This description serves as our single-particle part of our primary Hamiltonian. As seen in Fig.~\ref{fig:fig2}(a) and its inset, our ${\cal H}_{\rm band}$ features a hole pocket at the $\Gamma$-point from the valence band and an electron pocket at the $M$ point from the conduction band. As we shall see, such a topology of the bands helps us to understand the inter-band character of CDW ordering in TMDs. In addition to the above primary Hamiltonian, we also considered a secondary band structure, as depicted in Fig.~\ref{fig:fig2}(b), which presents a qualitatively different structure of the Fermi surface (FS). We will see later that such a dramatic change in the band structure between our primary and secondary models still results in qualitatively similar phase diagrams when interaction-induced correlated phases, such as CDW and SC, arise from ${\cal H}_{\rm int}$.

The above multi-band description of our toy model can be thought to originate from a microscopic tight-binding Hamiltonian defined on a triangular lattice 
of layered $1T$-\ce{TiSe_{2}}~\cite{PhysRevB.91.205135}, as schematically shown in Fig.~\ref{fig:fig2}(c). Here each lattice site contains two orbitals, $A$ and $B$, and the corresponding Hamiltonian takes the following form:
\begin{align}
    \label{eq:1}
	{\cal H}_0=&\sum_{i,j}t_{ij}^A(c_{Ai\sigma}^\dagger c_{Aj\sigma} + h.c.)
		+ \sum_{i,j}t_{ij}^B(c_{Ai\sigma}^\dagger c_{Aj\sigma} + h.c.)\nonumber\\
		&+ \sum_{i}t_p(c_{Ai\sigma}^\dagger c_{Bi\sigma} + h.c.)
\end{align}
Here, the first term represents the hopping of $A$-orbital electrons between sites $i$ and $j$ of a triangular lattice of size up to $N=96\times96$ with periodic boundary conditions. We consider that the electrons on orbital-A have finite amplitudes of hopping $t_A$ and $t_A^{\prime}$ between the nearest and next-nearest neighbors, respectively. In contrast, orbital-B allows only the nearest-neighbor hopping with amplitude $t_B$. The hopping of electrons between any other pairs of lattice sites is not considered in our minimal model. However, intra-orbital hopping of electrons with amplitude $t_p$ is considered only between the same lattice sites.
The schematic representation of the aforementioned tight-binding model is shown in Fig.~\ref{fig:fig2}(c). 
In momentum space, the above two-orbital tight-binding Hamiltonian in the basis
$\begin{pmatrix}c_{A{\bf k}\sigma}^\dagger&c_{B{\bf k}\sigma}^\dagger\end{pmatrix}$ can be expressed as:
\begin{align}
	{\cal H}_0({\bf k})=\begin{pmatrix}\epsilon^A_{\bf k}&t_p\\t_p&\epsilon^B_{\bf k}\end{pmatrix}.
\end{align}
Here,
\begin{align}
    \epsilon^A_{\bf k} &=\epsilon_A+2t_A(\cos(k_x)+2\cos(\frac{k_x}{2})\cos(\frac{\sqrt{3}k_y}{2})) \nonumber\\ 
    &+2t'(\cos(\sqrt{3}k_y)+2\cos(\frac{3k_x}{2})\cos(\frac{\sqrt{3}k_y}{2}))\\
    \epsilon^B_{\bf k} &=\epsilon_B+2t_B(\cos(k_x)+2\cos(\frac{k_x}{2})\cos(\frac{\sqrt{3}k_y}{2})).
\end{align}
Different combinations of parameters (e.g., $\epsilon_{A/B}$, $t_{A/B}$, $t'_A$, $t_p$) result in different band structures. Two sets of these parameters are responsible for the qualitatively different band structures in Fig.~\ref{fig:fig2}(a) and Fig.~\ref{fig:fig2}(b) See the caption of Fig.~\ref{fig:fig2} for the respective parameters. In fact, the diagonalization of the orbital Hamiltonian in Eq.~2 leads to the band Hamiltonian

\[{\cal H}_{\rm band}=\sum_{{\bf k},\sigma}\epsilon_{\bf k}^{\alpha} \alpha_{{\bf k}\sigma}^\dagger \alpha_{{\bf k}\sigma} + \epsilon_{\bf k}^{\beta} \beta_{{\bf k}\sigma}^\dagger \beta_{{\bf k}\sigma},\]
upon the basis change from orbitals $\{A,B\}$ to bands $\{\alpha,\beta\}$.


\subsection{Interactions}
\label{sec:sec2B}
Next, we include the effective electronic interactions in our toy model, which facilitate the correlated phases (CDW and SC) we wish to study. Experimental observations suggest that the charge-density wave state in $1T$-\ce{TiSe_{2}} is largely of excitonic origin~\cite{doi:10.1126/science.aam6432,Monney_2010,PhysRevB.103.125430}. Thus, the CDW does not arise from the usual Fermi-surface nesting within the same band or soft phonon modes, but rather from particle-hole pairing between distinct electronic bands, typically the \ce{Se}-4p valence band at the $\Gamma$ point and the Ti-3d conduction band at the $L$ point~\cite{PhysRevLett.99.027404}, and we will denote the strength of such inter-band interaction by $W$. In contrast, the SC being conventional~\cite{https://doi.org/10.1002/smll.202402749}, is assumed to be described by 
intra-band Hubbard attraction of strength, $U_{\alpha}$ and $U_{\beta}$. The pairwise interacting part of the Hamiltonian can therefore be written as: $\cal{H}_{\rm int}=H_{\rm CDW}+H_{\rm SC}$, where
\begin{align}
\label{eq:eq5}
    \cal{H}_{\rm CDW}&=W\sum_{\substack{{\bf k}, {\bf k}', {\bf q} \\ \sigma, \sigma'}} \alpha^{\dagger}_{{\bf k}+{\bf q}\sigma}\alpha_{{\bf k}\sigma} \beta^{\dagger}_{{\bf k}-{\bf q}\sigma'}\beta_{{\bf k}\sigma'}\\
    \cal{H}_{\rm SC}=&U_{\alpha}\sum_{{\bf k}{\bf k}'}\alpha^{\dagger}_{{\bf k}\uparrow}\alpha^{\dagger}_{-{\bf k}\downarrow} \alpha_{-{\bf k}'\downarrow}\alpha_{{\bf k}'\uparrow}\nonumber\\
    &+U_{\beta}\sum_{{\bf k}{\bf k}'} \beta^{\dagger}_{{\bf k}\uparrow}\beta^{\dagger}_{-{\bf k}\downarrow} \beta_{-{\bf k}'\downarrow}\beta_{{\bf k}'\uparrow}
\end{align}

The final Hamiltonian is therefore ${\cal H}={\cal H}_{\rm band}+{\cal H}_{\rm CDW}+ {\cal H}_{\rm SC}$, which we analyzed using mean-field theory
for a wide range of charge carrier density, $\rho$.

\subsection{Mean field analysis}

Employing a mean field decomposition of ${\cal H}_{\rm int}$ where ${\cal H}_{\rm CDW}$ produces the corresponding CDW ordering in the excitonic channel, and ${\cal H}_{\rm SC}$ generates conventional SC in the Bogoliubov channel, we arrive at the following mean field Hamiltonian,

\begin{align}
    {\cal H}_{\rm MF}&={\cal H}_{\rm const.}+\sum_{{\bf k}\sigma}\epsilon_{\bf k}^{\alpha} \alpha^\dagger_{{\bf k}\sigma}\alpha_{{\bf k}\sigma} + \sum_{{\bf k}\sigma}\epsilon_{\bf k}^{\beta} \beta^\dagger_{{\bf k}\sigma}\beta_{{\bf k}\sigma} \nonumber\\
    &-W\sum_{{\bf k}\sigma} (\chi_{\bf q}\beta^{\dagger}_{{\bf k}+{\bf q}\sigma}\alpha_{{\bf k}\sigma} + \chi_{\bf q}^*\alpha^{\dagger}_{{\bf k}+{\bf q}\sigma}\beta_{{\bf k}\sigma})\nonumber\\
    &+U_\alpha\sum_{\bf k}(\Delta_{\alpha}\alpha_{{\bf k}\uparrow}^{\dagger}\alpha^{\dagger}_{-{\bf k}\downarrow} + H.c.)\nonumber\\
    &+ U_\beta\sum_{\bf k}(\Delta_{\beta}\beta_{{\bf k}\uparrow}^{\dagger}\beta^{\dagger}_{-{\bf k}\downarrow} +  H.c.)
    \label{eq:eq7}
\end{align}

Here, $\chi_{\bf q}=N^{-1}\sum_{\bf k} \langle \alpha^{\dagger}_{{\bf k}+{\bf {\bf q}}\sigma}\beta_{{\bf k}\sigma}\rangle$ involves CDW pairing between electrons from the $\alpha$- and $\beta$-bands. The superconducting pairing amplitudes are given by $\Delta_{\alpha}=N^{-1}\sum_{\bf k}\langle \alpha_{{\bf k}\uparrow}^{\dagger}\alpha^{\dagger}_{-{\bf k}\downarrow}\rangle$ and $\Delta_{\beta}=N^{-1}\sum_{\bf k}\langle \beta_{{\bf k}\uparrow}^{\dagger}\beta^{\dagger}_{-{\bf k}\downarrow}\rangle$. Our numerical calculation involves iterative self-consistency of above conditions, we also need to find the chemical potential, $\mu$, self-consistently to fix the average electronic density, $\rho=N^{-1}\sum_{\bf k} (\langle \alpha^\dagger_{{\bf k}\sigma}\alpha_{{\bf k}\sigma}\rangle + \langle \beta^\dagger_{{\bf k}\sigma} \beta_{{\bf k}\sigma}\rangle)$, to the desired value. Note that in our system $\rho \in [0,4]$ because each ${\bf k}$-state can accommodate up to four electrons occupying either $\alpha$ or $\beta$ bands, and in each case, $\sigma=\uparrow$ or $\downarrow$. Having defined the model and methodology, we now discuss our key results.





\section{Results}
\label{sec:sec3}

\subsection{Phase diagram with charge modulation at nesting wave vector}
\label{sec:sec3A}

At half-filling $(\rho=2.0)$, inter-band nesting is strongest though imperfect, and charge modulation is expected to occur at the nesting vector ${\bf Q_{0}}$. Starting from this $\rho$, we crank up the density and investigate the fate of CDW order in the system; this amounts to mapping out the phase at $T=0$ as we discuss below. Here, ${\bf Q}_0$ is the nesting vector for the one-orbital model of the triangular lattice with nearest-neighbor hopping at the filling value of $\rho=1.5$. The Fermi surface is a hexagon with the vertex at the $M$-points. This geometry allows for three equivalent nesting vectors ${\bf Q}^{(1)}_0=(\pi, \pi/\sqrt{3})$, ${\bf Q}^{(2)}_0=(0,2\pi/\sqrt{3})$ and ${\bf Q}^{(3)}_0=(-\pi,\pi/\sqrt{3})$. In the case of two orbitals, the geometry changes but the Fermi surface symmetry remains intact. We will see that these are good choices for the ${\bf Q}$-vector in the pristine system. For our calculation we used ${\bf Q}^{(1)}_0$ alone, analogous to stripe phase, $1{\bf Q}$.

\begin{figure}[t]
    \centering
    \includegraphics[width=0.46\textwidth]{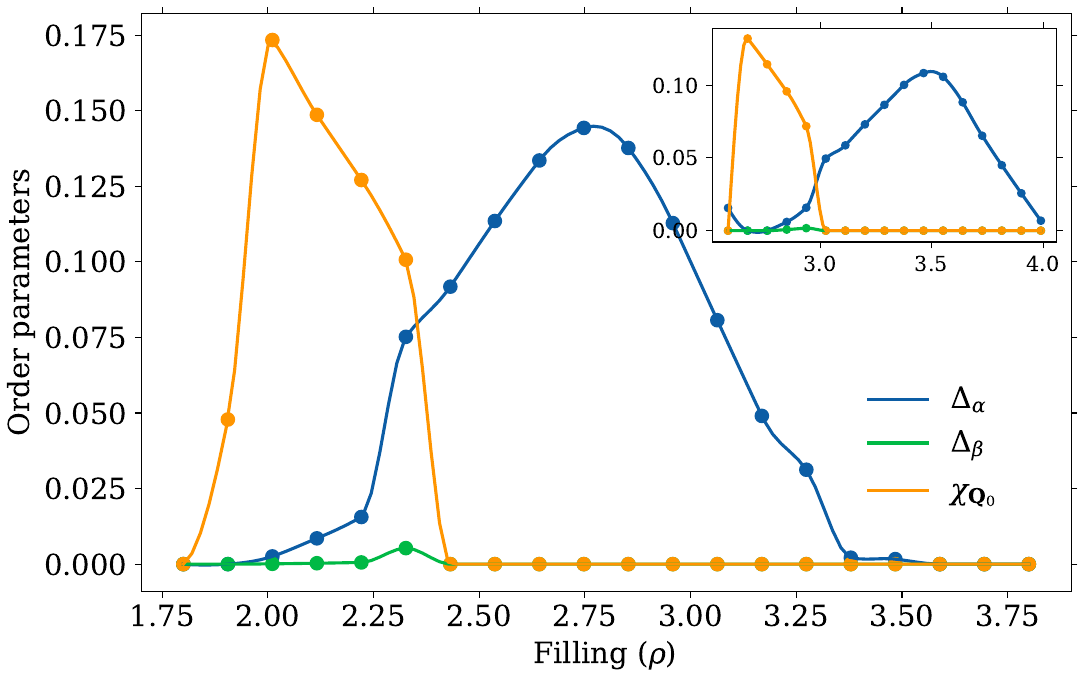}
    \caption{The phase diagram plotted for primary model with interaction strengths $W=8.0,U_\alpha=-1.1,U_\beta=0.0$ and for a fixed modulation vector of ${\bf Q_0}=(\pi,\pi/\sqrt{3})$. Inset shows the phase diagram for the secondary model with the following interaction values $W=3.625, U_\alpha=-1.1, U_\beta=0.0$.}
    \label{fig:fig3}
\end{figure}

The phase diagram corresponding to  ${\bf Q}^{(1)}_0$ is shown in the main panel of Fig.~\ref{fig:fig3} for our primary model, whereas a qualitatively similar phase diagram for the secondary model appears in the inset. The excitonic nature of CDW ensures that $\chi_{{\bf Q}_0}$ exists only in the window of the tuning parameter $\rho$, which corresponds to the overlapping region of the two bands. As we tune away, CDW degrades because the inter-band nesting weakens as the FS-topology changes. As we further tune the system, we are bordering on the region where the band overlap ends and deeper into the upper band; as a result, more and more states of the $\alpha$-band are now available to form Copper pair as a result SC is enhanced. This is observed in Fig.~\ref{fig:fig3} which is very reminiscent of the experimental phase diagrams of TMDs, particularly $1T$-\ce{TiSe_2}~\cite{Morosan2006,PhysRevB.110.165156,Li2016,PhysRevLett.103.236401}. In the phase diagram, we also find regions where both orders coexist. The calculation also confirmed that the generic nature of the phase diagram of Fig.~\ref{fig:fig3} remains robust to small changes to the band dispersion, i.e., insensitivity to the one-particle part ${\cal H}_0$ of the Hamiltonian. In addition, we carried out a position space calculation as outlined in Appendix.~\ref{sec:Appendix_B} which further verified that the phase diagram is robust to the basis in which interactions are cast, as shown in Fig.~\ref{fig:fig7}, indicating insensitivity to small tweaking of the interaction.
Our results confirm that the two orders compete to lift the degeneracy on the FS, and the dominant order is dependent on the hierarchy of the energy scales involved, that is $U_\alpha \Delta_\alpha$ and $W\chi_{{\bf Q_0}}$; for more details, refer to Appendix ~\ref{sec:Appendix_C}.

\subsection{Modification of the phase diagram from energy-minimizing CDW modulation}

\begin{figure}[t]
    \centering
    \includegraphics[width=0.5\textwidth]{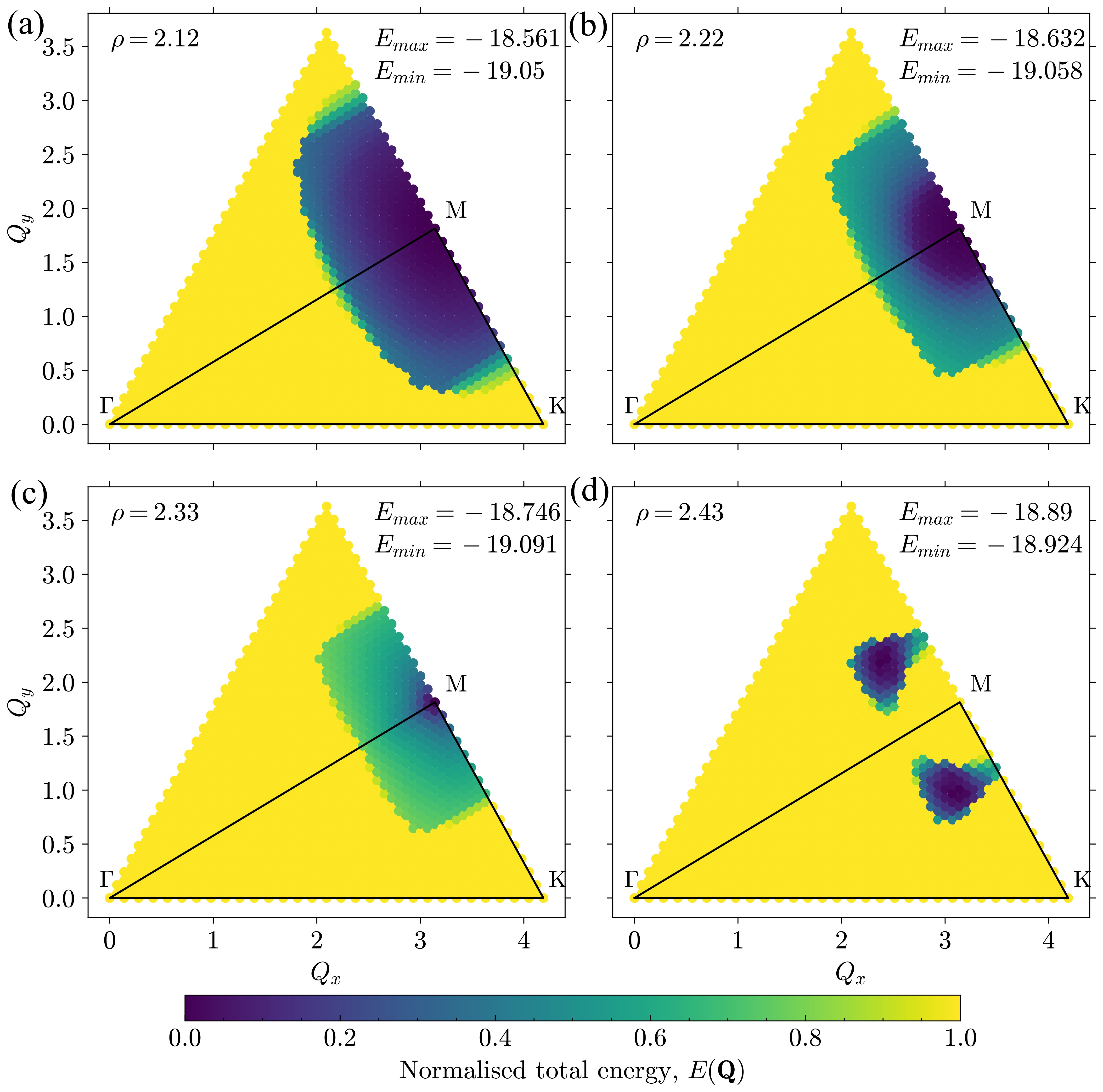}
    \caption{Normalized ground state energy $E({\bf Q})$ of the primary model with $W=8.0, U_1=-1.1$ and $U_2=0.0$ for all possible unique modulation wave-vectors chosen from the irreducible Brillouin zone. (a),(b) and (c) The global minima fixed at ${\bf Q}_0$ show the commensurate order. The energy compared to the normal-state energy ($E_{max}$) is weakened as the density is tuned. (d) Non-commensurate order: further tuning does not support the energy gain at ${\bf Q}_0$ and the minima split.}
    \label{fig:fig4} 
\end{figure}

In constructing the phase diagram in Fig.~\ref{fig:fig3}, we considered the charge ordering wave vector to remain fixed at the nesting vector of the underlying non-interacting problem at half-filling. On the other hand, we use the filling itself as the independent tuning parameter in the phase diagram. This description could become inconsistent because the Fermi surface topology changes with density, causing either the nesting vector to change or the nature of CDW to not remain nesting-mediated when $\rho$ is tuned. Away from half-filling, the interaction-mediated charge modulation can drift the ordering wave vector to a different value in the 2D $Q$-space. 
It is the minimization of energy corresponding to our Hamiltonian in Eq.~\ref{eq:eq7} which would determine the ordering wave vector of modulation for each $\rho$. Note that this process naturally selects the nesting vector as the energy-minimized charge ordering wave vector at $\rho=2$.

Motivated by this concept, we evaluate the ground state energy of our primary model for each $\rho$ as a function of the wave vector $({\bf Q})$, as follows:
\begin{align}
    E({\bf Q})=&\frac{1}{N}\sum_{{\bf k}}\sum_{i=1}^{8}E_{\bf k}^i f(E_{\bf k}^i)\nonumber\\
    &+ 2W\chi_{\bf Q}^2 - (U_\alpha \Delta_\alpha^2 + U_\beta \Delta_\beta^2)
\end{align}
Note that the first term in the above is the sum of the energies for all $k$-values in the first Brillouin zone, and the last two terms are the constant parts of ${\bf Q}$-dependent mean-field energies. We choose all possible values of the ${\bf Q}$-vectors from the irreducible Brillouin zone (IBZ)~\footnote{The irreducible Brillouin zone is the smallest unique k-space area due to the point group symmetries of the lattice and contains all the information needed to describe the electronic (or phononic) structure of the material} of the triangular lattice of a given size. We thus calculated $E({\bf Q})$ for all ${\bf Q}$ and selected the one as the charge ordering wave vector in the ground state that minimized $E({\bf Q})$.

The modified phase diagram upon incorporating the energy profile evolution is already shown in Fig.~\ref{fig:fig1} in the introduction. In fact, that figure contains the energy considerations outlined in Fig.~\ref{fig:fig4}. 
This result should be contrasted with the phase diagram in Fig.~\ref{fig:fig3} (main panel), where the CDW was assumed to originate from a fixed nesting vector at $\rho=2$ for all $\rho$. While the previous analysis remains valid deep within the band-overlap region--corresponding to Fig.~\ref{fig:fig4}(a)–(c), where the CDW energy scale dominates and is represented by a global minimum at the nesting vector--a distinct behavior emerges as the system approaches the boundaries of this region. In these marginal regions, the energy hierarchy is no longer dominated by a single CDW mode, and the system enters a crossover regime characterized by a small-amplitude CDW with a split modulation wave-vector $\mathbf{Q} = \mathbf{Q}_0 \pm \boldsymbol{\delta}$, as demonstrated in Fig.~\ref{fig:fig4}(d). This behavior is phenomenologically similar to a lock-in transition from a commensurate (C-CDW) to an incommensurate (I-CDW) phase~\cite{PhysRevB.14.1496,PhysRevB.12.1187}. It should be noted that while $\boldsymbol{\delta}$ typically represents a generic ${\bf k}$-space vector in physical ICDW transitions, our mean-field analysis constrains $\boldsymbol{\delta}$ to values compatible with the periodic boundary conditions (PBC) of the model. It is also important to note that we observe analogous physics also in the secondary model (For details, see Appendix~\ref{sec:Appendix_D}).

\subsection{Non-commensurate CDW at large $\rho$: Fermi surfaces picture}
\label{sec:sec3c}

\begin{figure}[t]
    \centering
    \includegraphics[width=0.5\textwidth]{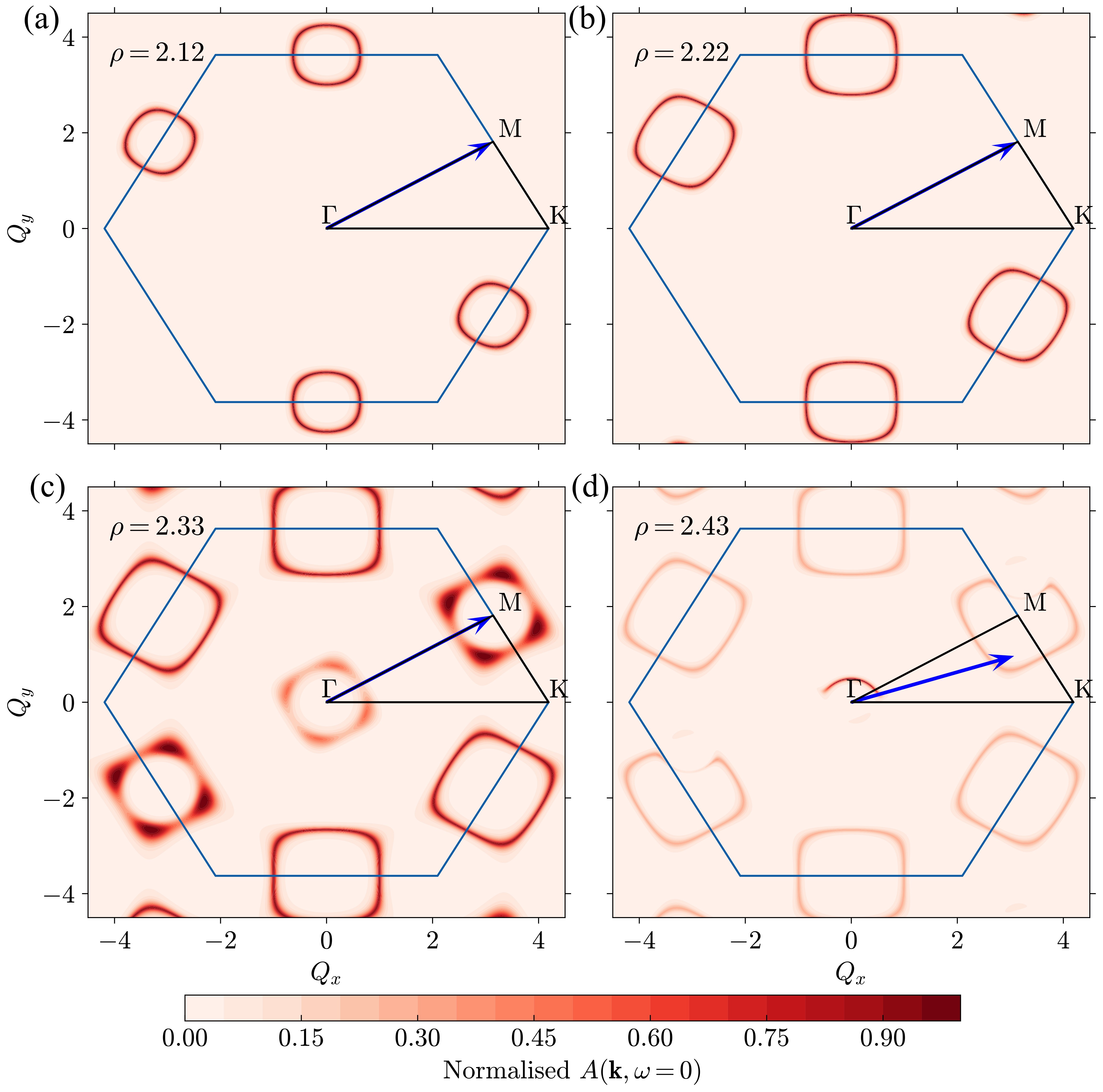}
    \caption{The spectral function corresponding to the FS $A({\bf k},\omega=0)$ for the ground state.(a)-(b) There is complete scattering between the electron pockets (at the $M$-point) and the hole pocket (at the $\Gamma$-point) also, the outer pocket has a slight increase in size (as seen from the unscattered part of the FS). (c) Due to enhanced filling, the FS at the $M$-point does not perfectly match the FS at the $\Gamma$-point, hence partial scattering. (d) Further tuning of density makes the scattering between the curvature of the outer FS and the inner circular points FS favorable.}    
    \label{fig:fig5} 
\end{figure}

To understand the origin of non-commensurate charge modulation at larger $\rho$ with a simultaneous emergence of SC, we next calculated the spectral function $A({\bf k}, \omega)$, from the retarded Green's function \( G^R({\bf k}, \omega) \):
\begin{eqnarray}
A({\bf k},\omega)&=&-\frac{1}{\pi}{\rm Im}[G^R ({\bf k},\omega)] \nonumber \\
&=&\sum_i (|V_{3i}|^2 + |V_{7i}|^2)d(\omega-E_{\bf k}^i),
\end{eqnarray}
where
\begin{equation}
G^R ({\bf k},\omega) =\sum_i  \frac{|V_{3i}|^2 + |V_{7i}|^2}{\omega-E_{\bf k}^i+i 0^+},
\end{equation}
and $V$ is a matrix whose columns are eigenvectors of the Hamiltonian in Eq.~(7). Only the third and seventh eigenvectors contribute because, in the quadratic form of ${\cal H}_{MF}$ written using Nambu-spinors, the third and seventh entries correspond to the electron part of bands $\alpha$ and $\beta$, respectively (For details, see appendix~\ref{sec:Appendix_A}). In Fig.~\ref{fig:fig5} we plotted \( A({\bf {\bf k}}, \omega = 0) \) over 1BZ. Here, $\omega=0$ represents the Fermi surface.

In Fig.~\ref{fig:fig5}(a), we presented  $A({\bf {\bf k}}, \omega = 0)$ near half-filling ($\rho=2.12$), for which the inter-band nesting dominates the quasi-particle scattering at ${\bf Q}_0$. As seen in Fig.~\ref{fig:fig5}(a), the FS near the $\Gamma$ and $M$ points of 1BZ is gapped out and validates this. As the electronic density increases to $\rho=2.22$, Fig.~\ref{fig:fig5}(b) shows that the electron pocket grows, but ${\bf Q}_0$ persists as the dominant scattering wave vector. At $\rho=2.33$ in panel (c), we see for the first time an incomplete scattering between hole-pocket at $\Gamma$-point and electron-pocket at $M$-point yielding partial gapping of the FS. Finally, for $\rho=2.43$ in Fig.~\ref{fig:fig5}(d), we see the emergence of a new scattering channel slightly away from the $M$-point, which causes the `splitting' of the charge modulation wave vector in Fig.~\ref{fig:fig4}(d), confirming non-commensuration.

 \subsection{Average density of states: A transition from one semi-metal to another semi-metal}

\begin{figure}[t]
    \centering
    \includegraphics[width=0.5\textwidth]{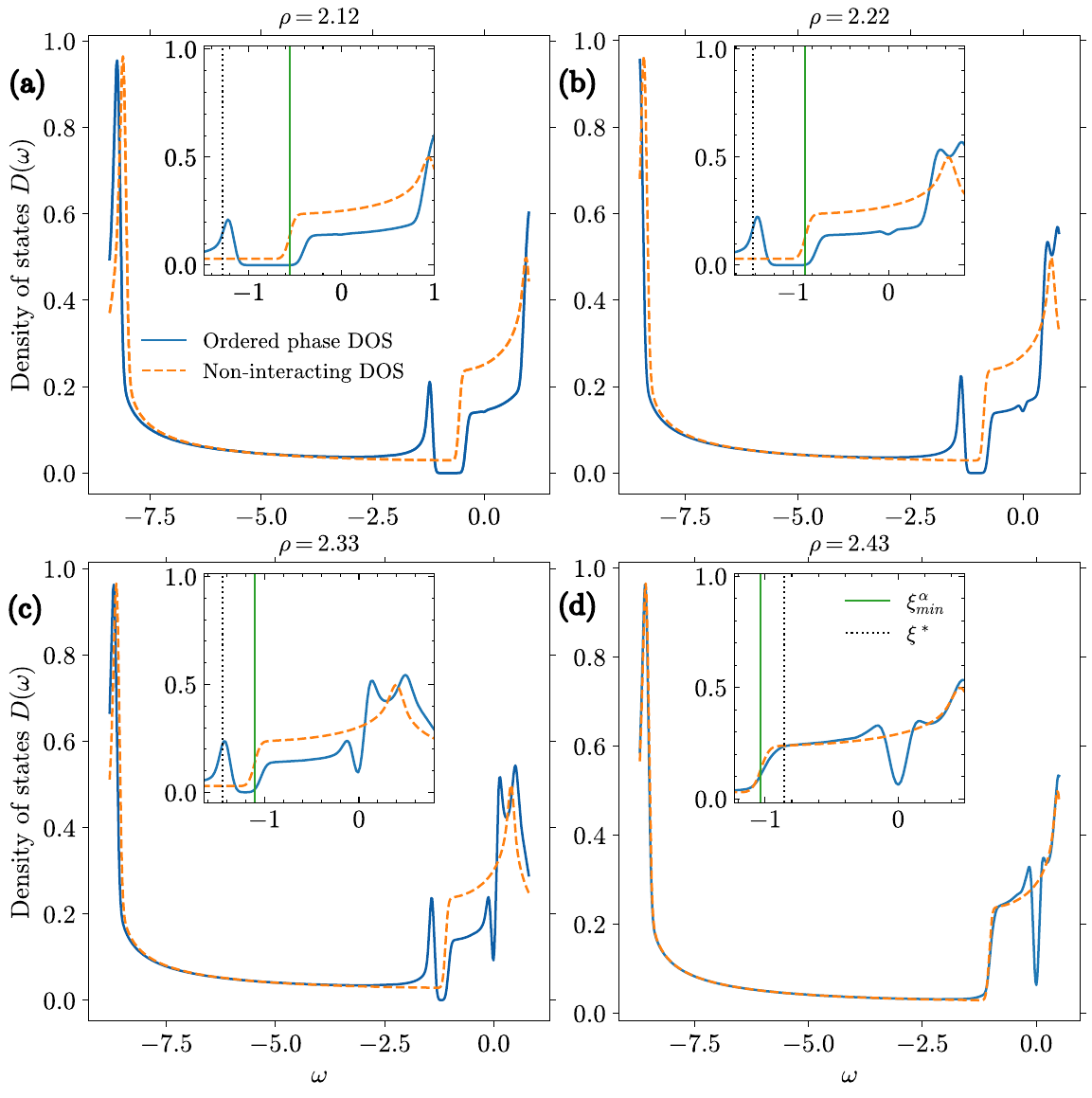}
    \caption{Density of States for corresponding fillings in Fig.~\ref{fig:fig4}. (a–b) A distinct CDW gap appears away from the Fermi level ($\omega=0$), pinned by the nesting energy $\epsilon^*$. (c) Coexistence regime showing both the CDW and SC kink at the Fermi level. (d) The CDW feature is suppressed as the SC gap dominates. Inset: Close-up near $\omega=0$, where $\xi_{min} = \epsilon^\alpha_{min} - \mu$ and $\xi^* = \epsilon^* - W\chi - \mu$.}
    \label{fig:fig6} 
\end{figure}

The changing nature of our system as $\rho$ is tuned can also be reflected in the local and average density of states -- an observable routinely tracked through scanning tunneling spectroscopy measurements~\cite{FEENSTRA1994965}. We present in Fig.~\ref{fig:fig6} the evolution of the average density of states (DOS), ${\cal N}(\omega) = N^{-1} \sum_{\bf k} A({\bf k}, \omega)$, for the ground state at various fillings $\rho$.
The DOS reveals a complex interplay of CDW and SC orders, a characteristic of TMDs. We observe that the CDW gap is pinned to the nesting energy, defined by the condition $\epsilon^{\alpha}({\bf k}) = \epsilon^{\beta}({\bf k} + {\bf Q_0})$ whose solution is considered as $\epsilon^*$,
in Fig.~\ref{fig:fig6}.
Consequently, the system avoids a metal-to-insulator transition. This behavior is in contrast to a standard one-dimensional Peierls transition~\cite{gruner1994density}, where the CDW order generates a global gap and drives the system into an insulating state. Because the nesting is not complete, it does not erase the FS entirely, leaving the CDW phase as a semi-metal, from the perspective of transport.
Interestingly, this picture is
qualitatively similar to the experimental findings on real $1T$-\ce{TiSe_2},
where it is believed that a semi-metal to semi-metal transition occurs~\cite{PhysRevLett.99.027404}.~However some Sulfur substitution experiments~\cite{PhysRevB.99.155103} were understood in terms semi-metal to semiconductor transitions. The distinction between these two is very thin, and our calculation is not equipped to settle this debate. 
It is worth mentioning that CDW and SC remain intertwined in our findings. Rather than only competing with each other, the presence of one reshapes the other -- here, CDW reshapes the DOS background over which the SC gap germinates.  
As the chemical potential changes upon tuning $\rho$, the spectral signatures of the CDW get buried. As a result, the opening of the SC gap at the Fermi surface becomes the dominant feature reflected in the phase diagram where SC enhances in the expense of the CDW phase. Thus, our results of ${\cal N}(\omega)$ illustrate the rich interplay of SC and CDW in TMDs.

\section{Discussion and conclusion}
\label{sec:sec4}

Starting with a minimal model of interaction-driven excitonic CDWs in a two-band triangular lattice, we unraveled its phase diagram, which captures the key experimental features. These include commensurate charge order in the pristine system, emergence of superconductivity with the collapse of CDW as the charge carrier density is increased, and finally, as SC emerges, the changing nature of charge modulation from commensurate to a non-commensurate ordering as SC emerges. The underlying physics can be understood by focusing on the evolution of FS topology with density and excitonic pairing. Experiments demonstrated an intriguing robustness of the phase diagram to various tuning parameters.~\cite{Morosan2006,PhysRevB.110.165156,Li2016,PhysRevLett.103.236401} In our calculations, we observe two levels of robustness. First, we find that the qualitative features of our phase diagram are insensitive to the details of the band structure (i.e., the one-particle part of the Hamiltonian). Second, the phase diagram is insensitive to the basis in which the interactions are cast; this is brought out by formulating an equivalent problem in position space (as outlined in Appendix.~\ref{sec:Appendix_B}) and comparing it to the momentum space result, which revealed a tantalizingly similar phase diagram, albeit having additional channels for interaction mediated charge modulation. These emphasize the robustness of the phase diagram to the tweaking of both single-particle and interacting terms of our toy model, which are broadly consistent with the experimental findings. There are other observations which our model does not address in its present form. These include, the absence of SC in Potassium deposition experiments~\cite{ZHANG2018426,PhysRevLett.130.226401}, effect of dimensionality~\cite{Chen2015,doi:10.1021/acsnano.5b06727,doi:10.1021/acs.nanolett.6b02710}, inhomogeneity~\cite{PhysRevB.92.081101,PhysRevLett.112.197001,PhysRevLett.118.106405,Lee2021PtTiSe2} and chirality~\cite{PhysRevLett.105.176401}, among others. We hope to extend our model to address these findings.

In conclusion, our minimal model successfully reproduces the qualitative phase diagram observed in experiments. This establishes that the interplay between CDW and SC can be understood through the lens of excitonic correlations and the evolution of Fermi surface topology with density. 

\section{appendix}
\subsection{Mean field Hamiltonian in quadratic form}
\label{sec:Appendix_A}
We can write the mean field Hamiltonian in the following way,
\begin{align}
    H_{MF}=\sum_{\bf k} \psi_{\bf k}^\dagger H_{\bf k} \psi_{\bf k} + H_{const}
\end{align}
Where,
\begin{align}
\psi_{\bf k}=\begin{pmatrix}
\alpha_{{\bf k}+{\bf q}\uparrow} & \alpha_{-{\bf k}-{\bf q}\downarrow}^\dagger & \alpha_{{\bf k}\uparrow} & \alpha_{-{\bf k}\downarrow}^\dagger & \beta_{{\bf k}+{\bf q}\uparrow} & \beta_{-{\bf k}-{\bf q}\downarrow}^\dagger & \beta_{{\bf k}\uparrow} & \beta_{-{\bf k}\downarrow}^\dagger
\end{pmatrix}^T
\end{align}
and in absence of any magnetic order spin rotational symmetry ensures that $\chi_{{\bf q}\uparrow}=\chi_{{\bf q}\downarrow}=\chi_{{\bf q}}$ and we have assumed that $\Delta_{\alpha,\beta}({\bf k}+{\bf q})=\Delta_{\alpha,\beta}({\bf k})=\Delta_{\alpha,\beta}$. hence we can write the $H_{\bf k}$ as,
\begin{align}
H = \begin{pmatrix}
\xi_{{\bf k}+{\bf q}}^\alpha & \Delta_\alpha & 0 & 0 & 0 & 0 & -\chi_{\bf q}^* & 0 \\
\Delta_\alpha^* & -\xi_{{\bf k}+{\bf q}}^\alpha & 0 & 0 & 0 & 0 & 0 & \chi_{\bf q} \\
0 & 0 & \xi_{{\bf k}}^\alpha & \Delta_\alpha & -\chi_{\bf q}^* & 0 & 0 & 0 \\
0 & 0 & \Delta_\alpha^* & -\xi_{{\bf k}}^\alpha & 0 & \chi_{\bf q} & 0 & 0 \\
0 & 0 & -\chi_{\bf q} & 0 & \xi_{{\bf k}+{\bf q}}^\beta & \Delta_\beta & 0 & 0 \\
0 & 0 & 0 & \chi_{\bf q}^* & \Delta_\beta^* & -\xi_{{\bf k}+{\bf q}}^\beta & 0 & 0 \\
-\chi_{\bf q} & 0 & 0 & 0 & 0 & 0 & \xi_{{\bf k}}^\beta & \Delta_\beta \\
0 & \chi_{\bf q}^* & 0 & 0 & 0 & 0 & \Delta_\beta^* & -\xi_{{\bf k}}^\beta \\
\end{pmatrix}.
\end{align}
note that the order parameters will be multiplied by respective interaction strengths $U_{\alpha,\beta}, W$ for SC and CDW, respectively. Also,
\begin{align}
    H_{const.}= 2NW\chi_{\bf Q}^2 - N(U_\alpha \Delta_\alpha^2 + U_\beta \Delta_\beta^2)
\end{align}
Then to diagonalize the Hamiltonian consider a unitary transformation $\psi_{\bf k}=U_{\bf k}\phi_{\bf k}$ such that $U^\dagger_{\bf k} H_{\bf k} U_{\bf k}=H_{\bf k}^D$ is diagonal. We can rewrite the order parameters and density in terms of the components of $U_{\bf k}$ and eigenvalues of $H_{\bf k}$. This way the values of the order parameters can be determined self-consistently.

\subsection{Comparison with a position-space calculation}
\label{sec:Appendix_B}

\begin{figure}[t]
    \centering
    \includegraphics[width=0.41\textwidth]{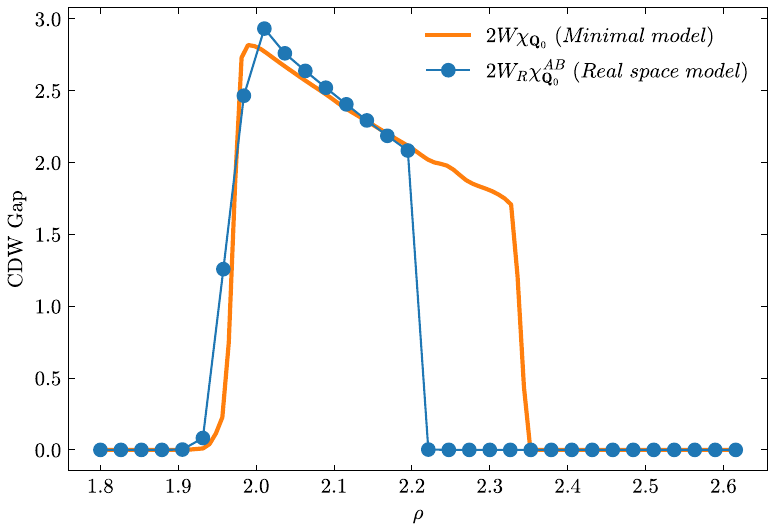}
    \caption{The evolution of the CDW gap with electronic density $\rho$ is presented from two independent calculations--one set up in position space, starting with a microscopic interaction defined in position space, and the momentum space analysis of Sec.~\ref{sec:2}. The high degree of similarity of the CDW phases from the two cases is evident. The parameters of the two calculations are set by matching the two solutions only at $\rho=2.14$. The modulation wave vector is fixed at ${\bf Q_0}=(\pi,\pi/\sqrt{3})$ and temperature is fixed at $T=0.01$. The minimal model only includes CDW with $W=8.0$. The position space model has the same tight-binding parameters as the minimal model with $W_R=8.0$. Note that with the above parameters, the conventional contribution to charge order is minimal (i.e. $\chi_{{\bf Q_0}}^A\approx0,\chi_{{\bf Q_0}}^B\approx0$).}
    \label{fig:fig7} 
\end{figure}


Phase diagrams are usually constrained by fundamental symmetries and symmetry breakings, and not by microscopic details, in general.
To establish the insensitivity of the topology of the phase diagram presented in Fig.~\ref{fig:fig3}, we used a different band structure representing our tight-binding Hamiltonian ${\cal H}_0$, as discussed in Sec.~\ref{sec:sec3A}. Here we address the effect of tweaking the interacting part of the Hamiltonian, ${\cal H}_{\rm int}$, and obtain the corresponding phases, restricting ourselves to the CDW-only part of the phase diagram.

To this end, we 
carried out a similar analysis described in Sec.~\ref{sec:2}; however, this time the mean-field analysis is set up in position space (unlike the momentum-space analysis presented in Sec.~\ref{sec:2}) starting with this microscopic Hamiltonian,
\begin{align}
	{\cal H}_{R}={\cal H }_{0}+W_R\sum_{i\sigma\sigma'}c_{Ai\sigma}^\dagger c_{Ai\sigma}c_{Bi\sigma}^\dagger c_{Bi\sigma}.
\end{align}
Here, ${\cal H }_{0}$ given in Eq.~\ref{eq:1}) is the tight-binding part defined on a triangular lattice of size $N_R=36\times36$ similar to that shown in Fig.~\ref{fig:fig2}(c) which, upon diagonalization, produces the bands with electron and hole pockets, depicted in Fig.~\ref{fig:fig2}(a). Following the same spirit of Coulomb interactions, here we consider a strong repulsive interaction between electrons of two different {\it orbitals,} denoted by $W_R$. This aspect is different from ${\cal H }_{\rm int}$ (Eq.~\ref{eq:eq5}) we considered earlier, where the interaction operated only among the electrons of distinct {\it bands}. The methodology to solve for the CDW phase arising from ${\cal H}_{R}$ is independent of that described in Sec.~\ref{sec:2}, and is based on mean field theory in real space similar to the Bogoliubov-de Gennes calculations for Superconductivity~\cite{Zhu2016BdG}. The Hamiltonian ${\cal H}_{R}$ takes the following form upon mean field decomposition in position space:
\begin{align}
    {\cal H}_{R} \stackrel{\text{MF}}{=} \sum_{ijl\sigma}&t^{l}_{ij}c_{li\sigma}^\dagger c_{lj\sigma}+\sum_{i\sigma}(t_p-W_R\Gamma_{Bi,Ai}^\sigma)c_{Ai\sigma}^\dagger c_{Bi\sigma}+H.c.\nonumber\\
    &+\sum_{il\sigma}(-\mu + W_R\sum_{\sigma'}\rho_{\bar{l}i\sigma'})c_{li\sigma}^\dagger c_{lj\sigma}.
\end{align}
Here, $t^{l}_{ij}$ ($l=A, B$) contains the information of intra-orbital hopping and onsite energies (i.e. $\epsilon_{A/B}$, $t_{A/B}$, $t'_A$) and $t_p$ is the inter-orbital hopping. We carried out a iterative self consistent calculation~\cite{GRT1998, GRT2001} for the fields defined as: $\rho_{li\sigma}=\langle c_{li\sigma}^\dagger c_{lj\sigma}\rangle$ and $\Gamma_{Ai,Bi}^\sigma=\langle c_{Ai\sigma}^\dagger c_{Bi\sigma}\rangle$ and chemical potential $\mu$ fixes the average electronic density defined as $\rho=N^{-1}_R\sum_{il\sigma}\rho_{li\sigma}$. The CDW is incorporated as a sinusoidal modulation of the self-consistent fields
\begin{align}
    &\rho_{li}=\rho_{l}^{(0)}+\chi^l_{\bf Q}e^{i{\bf Q}\cdot {\bf r}_i}\nonumber\\
    &\Gamma_{Ai,Bi}=\rho_{AB}^{(0)}+\chi^{AB}_{\bf Q}e^{i{\bf Q}\cdot {\bf r}_i}.
\end{align}

We choose the parameters of ${\cal H}_{R}$ such that, the resulting solution of $\chi^{AB}_{\bf Q}$ for $\rho=2.14$ matches with the results obtained in Sec.~\ref{sec:2} for the chosen nesting wave-vector ${\bf Q}_0=(\pi,\pi/\sqrt{3})$. Solving for $\chi^{AB}_{\bf Q}$ for other values of $\rho$, we present the results in Fig.~\ref{fig:fig4}, along with the corresponding results of the calculation scheme described in Sec.~\ref{sec:2}.
Interestingly, the CDW solution from the position-space calculation is predominantly inter-band (excitonic) in nature.
It is indeed reassuring that the phase diagrams from position- and momentum-space calculations with corresponding Hamiltonians are strikingly similar. 

\subsection{Interplay of CDW and SC: Phase diagram from different $U_1$ values}
\label{sec:Appendix_C}
\begin{figure}[b]
    \centering
    \includegraphics[width=0.5\textwidth]{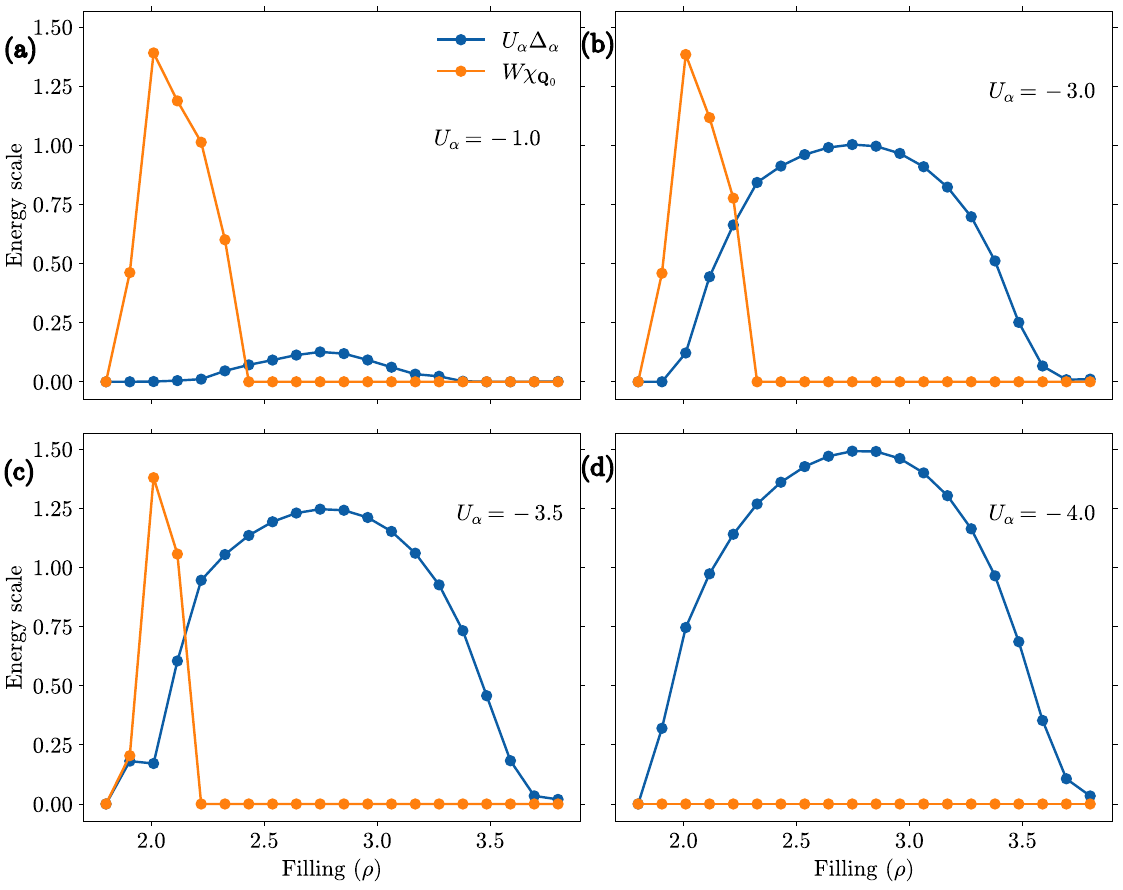}
    \caption{Phase diagram for ${\bf Q}=(\pi,\pi/\sqrt{3})$ and inter-band interaction strength of $W=8.0$ with different intra-band interaction (a) $U_\alpha=-1.0$ (b) $U_\alpha=-3.0$  (c) $U_\alpha=-3.5$  (d)  $U_\alpha=-4.0$. Note that $U_\beta=0.0$ for all the cases.}
    \label{fig:fig8} 
\end{figure}

Fig.~\ref{fig:fig8} illustrates the competitive interplay between CDW and SC orders as a function of their respective energy scales. Panel (a) corresponds most closely to our primary model, where the condition $W\chi_{\mathbf{Q}_0} \gg U\Delta_{\alpha}$ holds. In this regime, we observe asymmetric competition: the robust CDW order significantly suppresses the superconducting state while remaining relatively unaffected. As these energy scales become comparable in subsequent panels, the mutual exclusion intensifies, leading to a more pronounced intertwining of the two phases and a rapid suppression of the CDW order by the SC state.

\subsection{Splitting of minima in secondary model}
\label{sec:Appendix_D}
\begin{figure*}[t]
    \centering
    \includegraphics[width=1.0\textwidth]{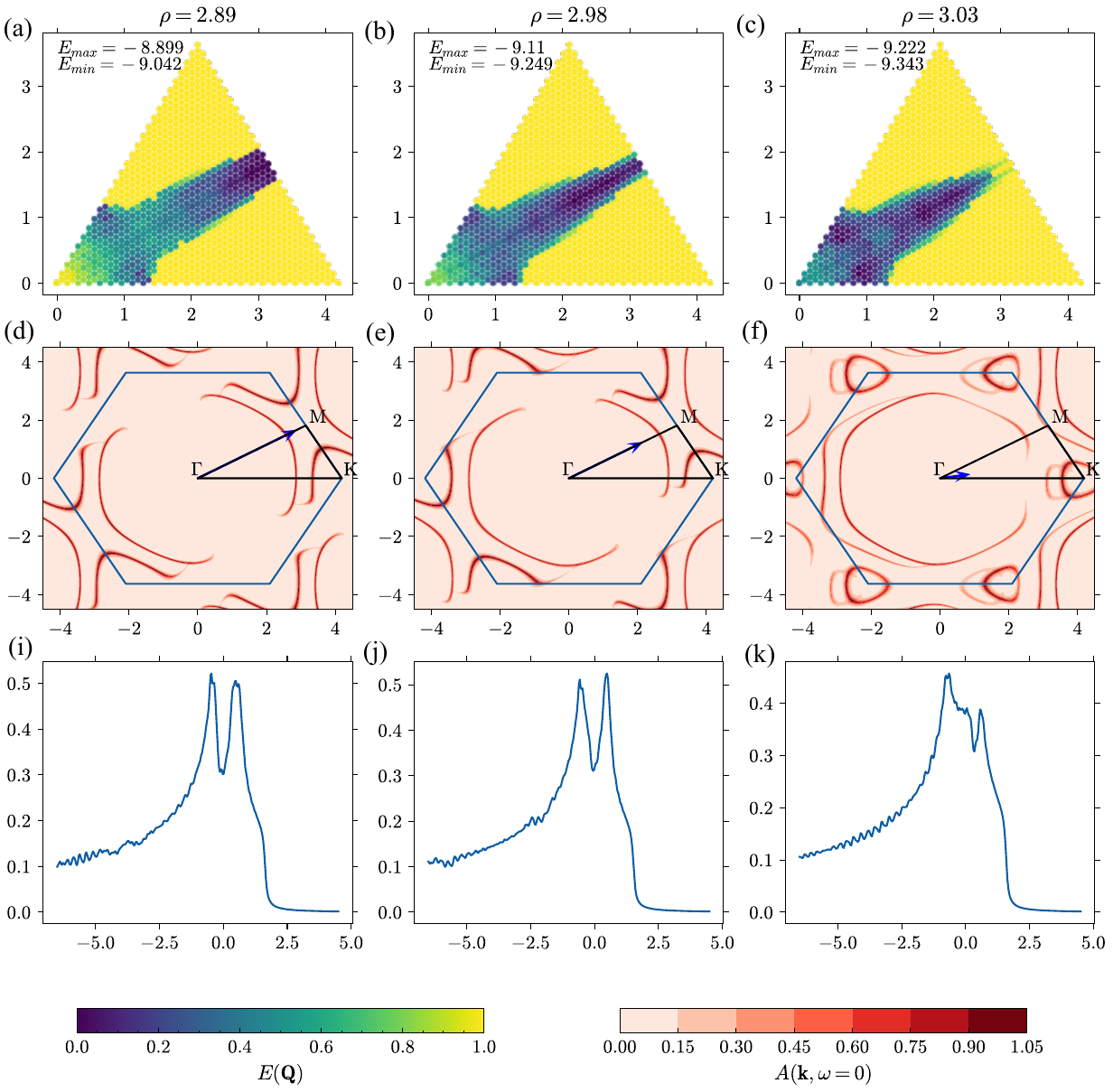}
    \caption{ (a),(b) and (c) Ground state energy calculated for $\rho=2.89$, $\rho=2.98$, and $\rho=3.04$, respectively. (d) (e) (f) Show the Fermi surface associated with the lowest-energy scattering for $\rho=2.89$, $\rho=2.98$, and $\rho=3.04$, respectively; gaped-out shows the scattering. (g)(h),(i) show the density of states corresponding to the lowest-energy scattering for $\rho=2.89$, $\rho=2.98$, and $\rho=3.04$, respectively.}
    \label{fig:fig9} 
\end{figure*}

We had observed the ``split" in the primary minimal model, which signified the onset of non-commensuration. In this section, we explored the qualitatively analogous physics of the secondary model. Here we have chosen three representative fillings from the phase shown in Fig.~\ref{fig:fig3} (inset). In Fig.~\ref{fig:fig9}(a) close to $\rho=2.89$, the ordering wave vector is close to the $M$ point, which is the nesting vector ${\bf Q}_0$. As we increase the filling to $\rho=2.98$, we see that the Q-vector has moved along the $\Gamma M$ line of ${\bf Q}$-space as shown in Fig.~\ref{fig:fig9}(b). It is to be noted that this still does not represent the ``split"; as we further increase the filling to $\rho=3.04$, we obtain two minima close to the $\Gamma K$ line as depicted in Fig.~\ref{fig:fig9}(c). The Fermi surface associated with the lowest-energy scattering demonstrates the underlying mechanism, i.e., the gaped-out and parallel regions from the inner band and the outer band show the scattering. Note that if we connect the parallel regions of the inner band and outer band with a scattering wave vector, it would lie outside the 1BZ; when folded back to the first Brillouin zone, it occurs close to the $M$-point as shown in Fig.~\ref{fig:fig9}(d). As we increase the total filling, the outer band will be more circular, and the inner-band FS will become more and more hexagonal with only curvature around the corners. Due to this, the lowest-energy scattering is now the ${\bf Q}$-vector connecting the arc of the outer FS with the rounder part at the corners of the hexagonal inner FS; hence the Q-vector is now at a tilt compared to ${\bf Q}_0$. When this tilted Q-vector is folded back to the 1BZ, it occurs at the point where we see the minima in Fig.~\ref{fig:fig9}(e). As the filling is further increased to $\rho=3.03$, the outer Fermi surface has been reduced to almost a circle; hence the modulation vector corresponds to scattering with the middle part of the edge of the inner hexagonal FS, which is what generates the split in Fig.~\ref{fig:fig9}(f). The DOS corresponding to the lowest-energy scattering reveals that the system retains its metallic nature throughout the cross-over due to the incomplete scattering akin to the primary model.

\FloatBarrier
\bibliographystyle{apsrev4-2}
\bibliography{references}

\end{document}